\documentclass[prl,twocolumn]{revtex4}
\def\BibTeX{{\rm B\kern-.05em{\sc i\kern-.025em b}\kern-.T
    08em\kern-.1667em\lower.7ex\hbox{E}\kern-.125emX}}
\usepackage{graphicx}

\begin{document}

\title{Long-distance Transfer and Routing of Static Magnetic Fields}

\author{C. Navau$^{1}$}
\author{J. Prat-Camps$^{1}$}
\author{O. Romero-Isart$^{2,3}$}
\author{J. I. Cirac$^4$}
\author{A. Sanchez$^{1,*}$}

\affiliation{$^1$Departament de F\'isica, Universitat Aut\`onoma de Barcelona, 08193 Bellaterra, Barcelona, Catalonia, Spain}
\affiliation{$^{2}$Institute for Quantum Optics and Quantum Information of the
Austrian Academy of Sciences, A-6020 Innsbruck, Austria}
\affiliation{$^{3}$Institute for Theoretical Physics, University of Innsbruck, A-6020 Innsbruck, Austria}
\affiliation{$^4$Max-Planck-Institut f\"ur Quantenoptik,
Hans-Kopfermann-Strasse 1,
D-85748, Garching, Germany.}


\begin{abstract}
We show how the static magnetic field of a finite source can be transferred and routed to arbitrary long distances. This is achieved by using transformation optics, which results in a device made of a material with a highly anisotropic magnetic permeability. We show that a simplified version of the device, made by a superconducting-ferromagnet hybrid, also leads to  an excellent transfer of the magnetic field. The latter is demonstrated with a proof-of-principle experiment where a ferromagnet tube coated with a superconductor improves the transfer of static magnetic fields with respect to conventional methods by a 400\% factor over distances of 14cm.



\end{abstract}

\maketitle

Magnetism is a fundamental interaction shaping our physical world, at the basis of
technologies such as magnetic recording or energy generation. 
Guiding and transferring magnetic fields is essential in many technologies, from large scale transformers to nanoscale magnetic logic devices \cite{coey,carlton}.
Unlike electromagnetic waves, which can be
routed and transmitted with waveguides or optical fibers to long distances, static magnetic fields
rapidly decay~\cite{Jackson}. The conventional way to transfer magnetic fields is using a ferromagnetic (FM) material with high magnetic permeability $\mu$, as in the transfer of magnetic field from the primary to the secondary circuits of a transformer~\cite{coey}.
However, the transferred field rapidly decays with distance. We show in Fig. 1 a demonstration of this fact for a cylindrical geometry, which will be employed throughout the work. The reason for this is that the isotropic $\mu$ yields high values of magnetic induction {\bf B} not only in the direction of the cylinder axis but also in the radial direction, so field escapes through the cylinder lateral surface and the value at its end decreases drastically with increasing the length.
Alternatively, one could consider to guide magnetic field lines through a hollow superconducting (SC) tube. However, the transferred field rapidly decreases with increasing the tube length
(see Fig. 1 and \cite{levin}). Hence, in contrast to time-dependent electromagnetic fields, which can be transmitted and routed to long distances with e.g. optical fibers, a device capable to do so with static magnetic fields does not exist.

In this Letter we theoretically design and experimentally demonstrate such a device, which we call \emph{magnetic hose}. 
The design of the magnetic hose is based on transformation optics~\cite{TOpendry,review_TO,controlling,TOpendry_science},
which has already enabled realizing invisibility cloaks \cite{controlling,microwave,chen_alu}, perfect lenses 
\cite{perfectlens}, and waveguides operating in the subwavelength regime
\cite{zhang}.  Transformation optics has been succesfully applied also to static fields \cite{wood,Magnus2008,antimagnet,narayana,gomory_science,dccloak,dcconcentrator,concentrator} and has allowed to experimentally implement magnetic cloaks~\cite{narayana,gomory_science}. The application of transformation optics to static magnetic fields allows us to find that an infinitely wide slab of thickness $t$ made of a homogenous material with $\mu_{\parallel}=\infty$ along the thickness, and $\mu_{\perp}=0$ along the width, exactly transfers the magnetic field from the lower surface of the slab to the upper surface (see Figs. 2a and 2b for the case of a dipolar source and also Supplemental Material Sect. I). This material effectively transforms the space such that the region below the slab is unaltered and the space within the slab is shifted to the region above it. Such a slab has other remarkable properties. First, magnetic induction {\bf B} is vertical and magnetic field {\bf H} horizontal in the material volume. Because {\bf B } is perpendicular to {\bf H}, the magnetic energy is zero in the material \cite{concentrator}. Second, {\bf B} and {\bf H} -and thus magnetic energy- are not modified in the region containing the sources. The magnetic energy originally in the material volume is thus transferred to the other end. This is valid for any magnetic source and, remarkably, for any slab thickness $t$. However, this ideal design is not feasible, both because (i) the slab is infinitely wide and because (ii) no materials exist with such extreme anisotropy.


\begin{figure}[t]

\centering

\includegraphics[width=0.45\textwidth]{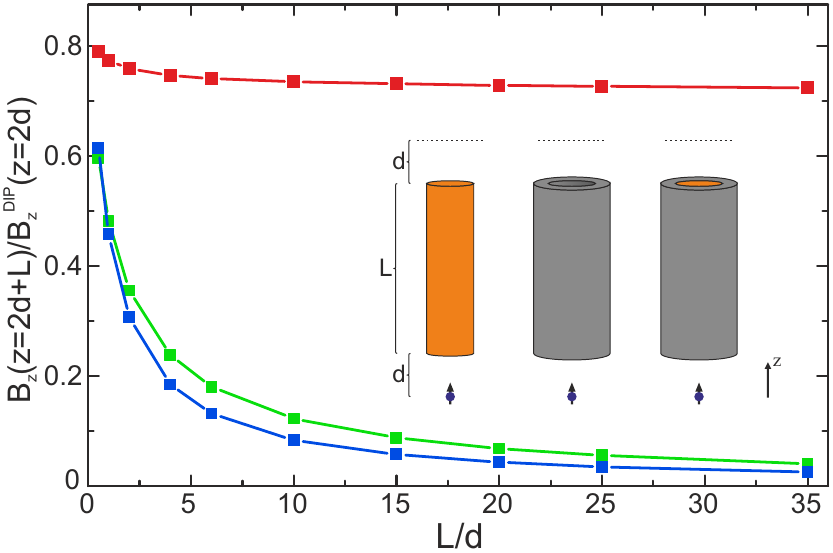}

\caption{The maximum field at a distance 
$d$ of the top surface $B_z(z=2d+L)$, normalized to the maximum field of 
the isolated dipole at a distance $2d$, $B_z^{\rm DIP}(z=2d)$,  is plotted for an ideal FM 
cylinder (green line), an ideal SC tube (blue line) and a bilayer magnetic hose 
(red line), when a dipolar source is placed at a distance $d$ below them 
as a function of the cylinder lengths $L$. In the inset, sketches of the three cases are shown, with the FM in orange and the SC in gray.
In all these calculations the radius of the FM cylinder is $r=0.8d$ and 
the thickness of the SC tube $w=0.5d$.}

\end{figure}


\begin{figure}[t]

\centering

\includegraphics[width=0.45\textwidth]{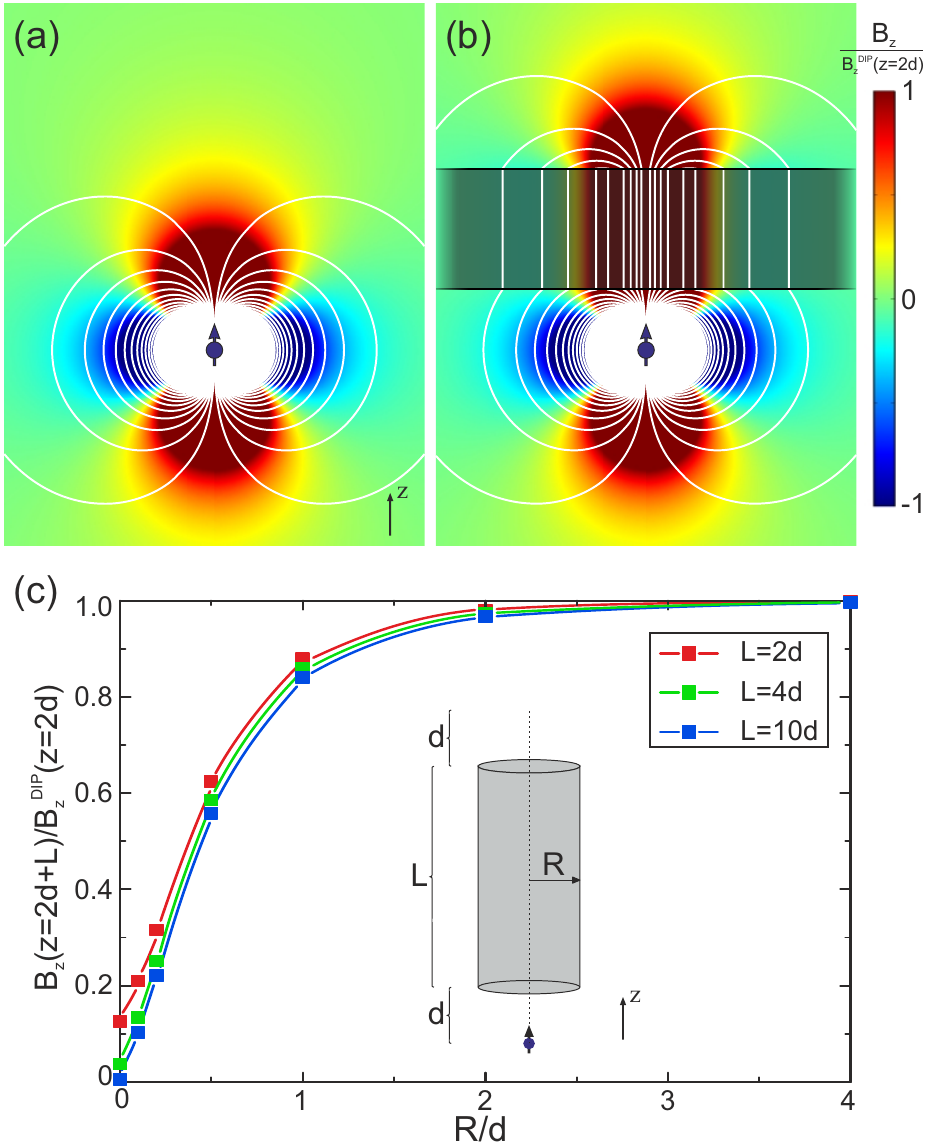}

\caption{The field of a magnetic source, e.g. a point dipole (a) (field 
lines in white and vertical field $B_z$ in colors), is exactly 
transferred to an arbitrary distance, (b), through an infinite slab of 
ideal material (shaded region) having $\mu_{\parallel}=\infty$ along the thickness and $\mu_{\perp}=0$ along 
the width. (c) Maximum vertical field at a distance $d$ above a magnetic hose made of the ideal material, when a vertical dipole is placed at a distance $d$ below the hose, as a function of the radius $R$. The field is normalized to the maximum field created by an isolated dipole at 
a vertical distance $2d$, $B_z^{\rm DIP}(z=2d)$.  Different hoses of length $L=2d$, $4d$ and $10d$ are plotted in red, green, and blue, respectively. }

\end{figure}


\begin{figure}[t]

\centering

\includegraphics[width=0.45\textwidth]{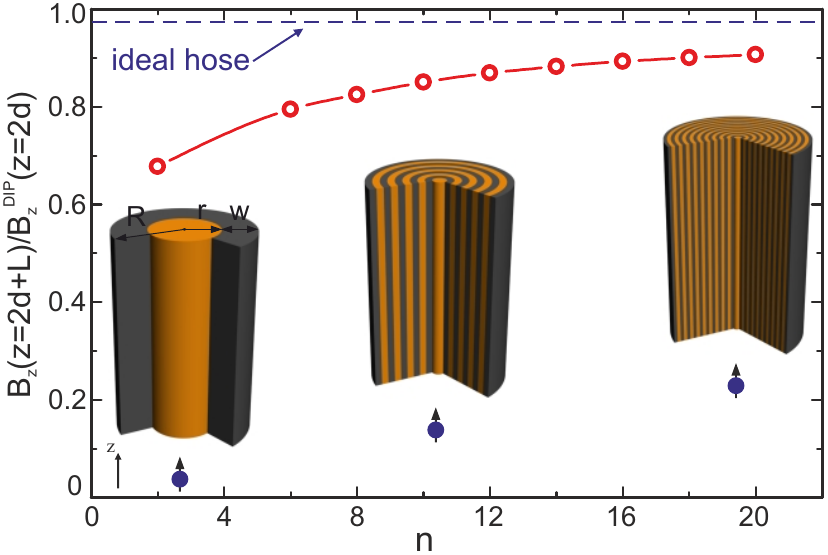}

\caption{Maximum field transferred by hoses discretized into different 
number of layers $n$. A point dipole is placed at a distance $d$ below 
the bottom end and the field is calculated at a distance $d$ above the 
top. SC and FM layers have all the same thickness, as shown in the 
insets (FM in orange and SC in gray). The field 
transferred by an ideal hose with the same geometry is plotted with a blue dashed line.
The total radius of the hose is $R=2d$ and its length $L=4d$.}

\end{figure}


\begin{figure}[t]

\centering

\includegraphics[width=0.45\textwidth]{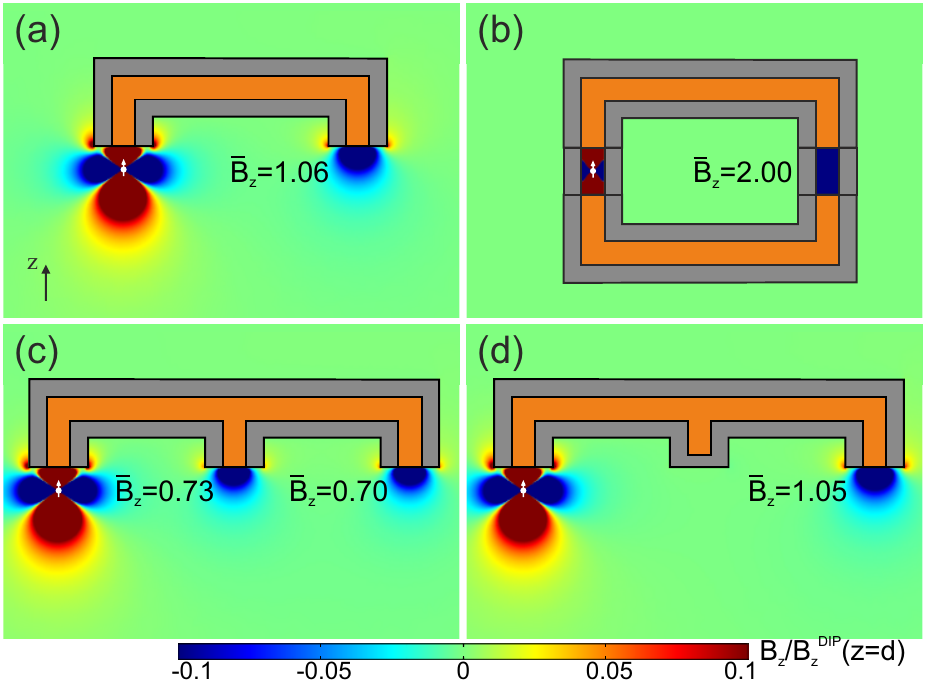}

\caption{Vertical field component $B_z$ (in color scale) in the middle 
plane of different 3D geometries. The field source is a point dipole 
(sketched in white) placed at a vertical distance $d$ from the hoses. 
The FM cylindrical core (orange) has a radius $r=0.5d$ and the 
surrounding SC (gray) a thickness $w=0.75d$.
In each simulation the average output field $\bar{B}_z$ is calculated as 
the quotient between the flux that exits each hose end and the area of 
the ferromagnetic section and normalized to the maximum vertical field 
created by an isolated dipole at a distance $d$, $B_z^{\rm DIP}(z=d)$. 
Analytical and numerical values of $\bar{B}_z$ fully agree for the 
closed hose (b).
Case (c) shows a hose divided into two branches; when one of its ends is 
closed with the SC, (d), we recover the case of a hose with a single 
branch (a).
See Supplemental Material Sect. VI for further discussions of the results.}

\end{figure}


\begin{figure}[t]

\centering

\includegraphics[width=0.45\textwidth]{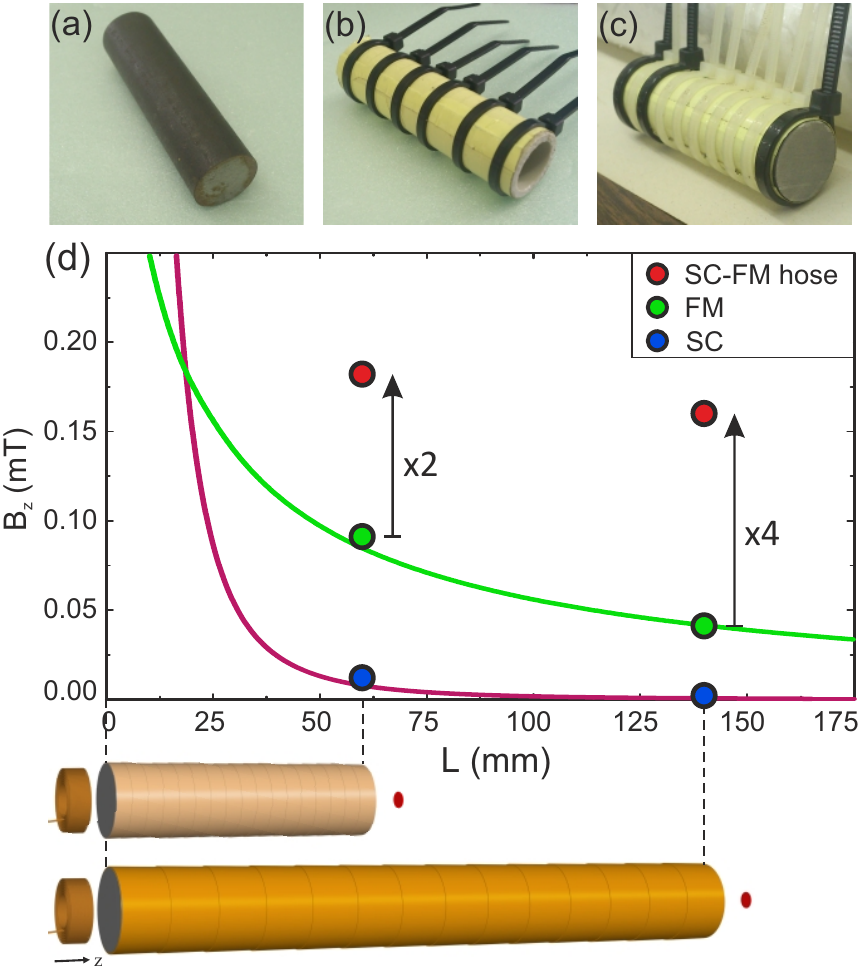}

\caption{Pictures of the construction of the experimental devices (of 
length 60mm); (a) the FM cylinder (b) the SC shell wrapped around a 
plastic former and (c) the SC+FM bilayer magnetic hose. (d) Measured transferred magnetic 
field at one end of cylinders of length 60mm and 140mm. At the other end 
the field source is a coil fed with 1A dc current (see the sketches 
below). Three cases are measured: SC only, FM only, and the SC+FM bilayer magnetic hose. 
Lines are calculated results for the field of the bare coil (purple 
curve) and that transferred by the FM only (green).}

\end{figure}

Let us address these difficulties separately. To circumvent (i), we consider a cylindrical piece of material with finite radius $R$, length $L$,
radial permeability $\mu_{\rho}=0$, and axial permeability $\mu_{z}=\infty$. We numerically study the transfer of magnetic field when a vertically-aligned magnetic dipole is placed at a distance $d$ below its bottom end (inset in Fig. 2c). In Fig. 2c we show the maximum vertical field $B_z$ (always found at $\rho=0$) at a distance $d$ above the top end as a function of $L$ and $R$. $B_z$ decreases only slightly when the material length increases and rapidly saturates to a certain value that depends on $R$. It can be demonstrated (based on magnetic poles, see Supplemental Material Sect. II) that this saturated value is not zero, hence, a part of the magnetic field is always transferred through the hose to arbitrary distances. For $R=4d$ and lengths up to $L=10d$ the material transfers the field from one end of the hose to the other, similarly to the ideal infinite slab. This property is approximately maintained even for radii as small as $R=2d$.

To circumvent difficulty (ii) we follow a similar strategy used in the context of magnetic cloaks and concentrators \cite{antimagnet,concentrator}. The extreme anisotropic medium with $\mu_{\rho}=0$ and $\mu_{z}=\infty$ can be approximated 
using existing isotropic materials: a series of $n$ alternated superconducting (ideally with $\mathbf{B}=0$ in their volume) and soft ferromagnetic (ideally with $\mathbf{H}=0$) concentric cylindrical layers (see Supplemental Material Sect. VIII for a discussion on the adequacy of the materials). Ferromagnetic parts provide a large axial permeability and the alternated superconductors prevent radial {\bf B} components.
In Fig. 3 we show calculated results for several discretized versions of the magnetic hose into $n$ concentric cylindrical shells, half of them superconducting and half ferromagnetic.  The field of the dipole, located at one end of the hose, that is transferred to the opposite end tends to the ideal behavior with increasing $n$. More than 90\% of the transfer can be achieved with $n$=20. Remarkably, even with the simplest configuration ($n$=2) the transfer can be as high as 75\% (as in Fig. 1). In this bilayer scheme the SC has to be the outer layer, surrounding the FM, in order to prevent the field lines attracted by the inner FM core to leak outside through the lateral surface. 

We concentrate hereafter on the simplest hose discretization, a SC shell of thickness $w$ surrounding a FM core of radius $r$ ($R=r+w$). 
The field transmitted by this simple bilayer hose strongly depends on its sizes $r$ and $w$, because they determine the amount of field lines that go into the hose. In section IV of the Supplemental Material we show that an optimum transmission is obtained with $r=0.6d$ and $w= 1.4d$.
In Figs. 4 we show numerical calculations illustrating the routing of the magnetic field of a dipolar source through hoses with different topologies with open and closed exits; many combination of sources, branching geometries, and open and closed ends can be made following these examples. Results provide a clear verification of the capabilities of the magnetic hose -even in its simplest discretization- for routing and long-distance transfer of magnetic fields (see Supplemental Material Sect. VI for further examples). As mentioned above, we emphasize that this cannot be obtained with only the FM or the SC (see a comparison in Fig. 1).

For the case of closed hoses (as in Fig. 4b), we have validated the numerical results, from first principles, by analytically solving magnetostatic Maxwell equations with the boundary conditions: {\bf B} is zero in the SC and the scalar magnetic potential is constant in the FM. As shown in the Supplemental Material Sect. V, the field of the dipolar source with magnetic moment $m$ in the left branch is transferred towards the air gap (right), and builds up there according to the expression
\begin{equation}
B_z=\frac{\mu_0 m}{2\pi r^2 (l_{\rm s}+l_{\rm a})}.
\end{equation}
$r$ is the radius of the FM core,  $\mu_0$ the vacuum permeability, and $l_{\rm s}$ ($l_{\rm a}$) the length of the source (air) gap. Eq. (1) shows that the field in the whole air gap is uniform and vertical \cite{cioffi} and, remarkably, does not depend on the hose length. 

We have experimentally confirmed the magnetic field transferred through bilayer SC-FM magnetic hoses. 
We constructed two hoses with lengths 60mm and 140mm, each consisting of a FM cylinder and a SC shell made by adjacent pieces of YBCO superconducting tape 
(see Supplemental Material Sect. VII). A constant current loop was placed at one end of the hoses and the transferred magnetic field was measured at the opposite end by a Hall probe. Results in Fig. 5a show the measured transferred field for the SC only, the FM only, and the magnetic hose. We also show the calculated transferred field by the FM only (green curve), showing excellent agreement with the experiments, and the field of the the bare coil (purple). Field transferred along the hose is much larger than that with only the FM or the SC and the ratio between the fields transferred by the hose and the FM increases with increasing length, confirming our theoretical ideas.  The field transferred by the hose is less than the theoretical value and slightly decreases with increasing length because a non-perfect sealing of SC parts which can be improved by a better manufacturing. This is supported by numerical calculations which agree with the experimental results, see Supplemental Material Sect.VII.
Even with these imperfections in such a proof-of-principle experiment, the measured transferred field ratio of the hose with respect to the FM increases from 200\% to 400\% between the short and long studied lengths.

Since the magnetic hose operates with static fields, associated with an infinite wavelength, our ideas can be implemented at any scale
\cite{wood,gomory_science,concentrator,TOpendry_science}. This hints at the possibility of miniaturizing the magnetic hose
to the nanoscale. Such a magnetic nanohose could open new possibilities to increase the magnetic coupling between nanomagnets
in magnetic logic schemes, which has been identified as an important issue in some spintronic proposals \cite{carlton,cowburn_Science,cowburn_NJP,science_maglogic}.
A magnetic nanohose could also be used as a new tool to harness quantum systems, as required, for instance,
in quantum information processing~\cite{SIQI}. In particular, such a device would permit to address, control, and manipulate the internal
state of individual quantum systems, even if they are separated by distances of the order of few tens of nanometers,
where optical methods are no longer available. This could be particularly relevant in the context of Nitrogen-vacancy (NV) color
defect centers in a diamond nanocrystal~\cite{NV}, identified as promising systems for the implementation
of quantum information processors \cite{Norm}, or quantum repeaters \cite{repeaters}. Beyond the transfer of classical magnetic fields, one 
could also envision a similar system as the one
considered here to couple distant quantum systems magnetically, allowing to separate them by relatively long distances and still
strongly interact with each other. This would provide us with a very powerful alternative to existing methods that directly
couple NV centers at the quantum level \cite{natphys}. 
The characterization of such a system, however, requires a full quantum treatment, something that is
out of the scope of the present work.

In conclusion, we have theoretically proposed and experimentally demonstrated a magnetic hose to transfer the static magnetic field of a given source to distant positions along chosen paths. The hose is designed to operate with static magnetic fields and it is the actual magnetostatic field lines that are guided. We remark that other approaches recently used for manipulating and guiding electromagnetic waves, even at subwavelength scales, such as epsilon-near zero (ENZ) materials \cite{silveirinha,liu,ENZ} or hyperbolic metamaterials \cite{hyperbolic}, cannot be used with static fields. The reason is that these approaches are based on the coupling of electric and magnetic fields, whereas in our static case the displacement current coupling electric and magnetic field is negligible. Moreover, as discussed in \cite{wood}, ENZ metamaterials at zero frequency do not exist. The magnetic hose proposed here thus brings a new tool to control magnetic fields, which complements the possibilities offered by metamaterials for the case of electromagnetic waves.

We thank Peter Zoller and John Clem for comments, Xavier Granados for support, and Vacuumschmelze and Superpower for providing the ferromagnetic cylinder and the superconducting tape, respectively. We acknowledge funding from the EU projects AQUTE, and Spanish Consolider NANOSELECT (CSD2007-00041) and MAT2012-35370 projects. AS acknowledges funding from ICREA Academia, Generalitat de Catalunya. JPC acknowledges a FPU grant form Spanish Government (AP2010-2556). ORI acknowledges funding from ERC-StG QSuperMag.

(*) Corresponding author: alvar.sanchez@uab.cat.

\end{document}